\documentclass{PoS}

\title{BSM Lessons from the SM Higgs}

\ShortTitle{BSM Lessons from the SM Higgs}

\author{\speaker{Andrea Wulzer}%\thanks{A footnote may follow.}
\\
        Dipartimento di Fisica e Astronomia ``G.~Galilei'', Universit\`a di Padova
and INFN, Sezione di Padova, Via Marzolo 8, I-35131 Padova, Italy\\
        E-mail: \email{andrea.wulzer@pd.infn.it}}

\abstract{In this talk I will review the implications on Standard Model (SM) and Beyond the SM (BSM) theory of the experimental exploration of the scalar sector. Given that the Higgs discovery has been the most important achievement, I will start with a general overview of its dramatic impact on the physics of fundamental interaction. Next, I will describe how scalar sector results, including the measurements of the Higgs couplings and mass and the negative searches for extra scalars, constrain concrete BSM scenarios related with Naturalness. In particular, I will discuss  composite Higgs and low energy supersymmetry, both in its minimal and next-to-minimal implementations.}

\FullConference{The European Physical Society Conference on High Energy Physics\\
		22--29 July 2015\\
		Vienna, Austria}

\begin{document}

\section{Why the Higgs is Revolutionary}

The discovery of the Higgs boson \cite{hdisc-atlas,hdisc-cms}, with properties compatible with the Standard Model (SM) expectations \cite{Aad:2015gba,Khachatryan:2014jba}, is sometimes erroneously regarded as a ``boring'' confirmation of the SM theory. On the contrary, it is a revolutionary event in the history of fundamental interactions physics. It is of course a confirmation of the SM, but not at all a boring one. I will describe the dramatic implications of the SM Higgs discovery in the next section, through the concept of ``No-Lose Theorems''.

\subsection*{No-Lose Theorems}

A No-Lose Theorem is an incontrovertible theoretical argument, purely based on currently experimentally established facts, that \emph{guarantees future discoveries}, to be achieved provided the experimental conditions become favourable enough. In the concrete examples discussed below,  ``favourable enough'' conditions simply means an high enough collider energy. The No-Lose Theorems that follow, each at its historical moment, thus offered absolute guarantees of new physics discoveries through the study of higher and higher energy collisions, providing  unquestionable motivations for the exploration of the energy frontier at more and more energetic colliders.  After the Higgs was discovered and the SM was found to be a valid description of the Electro-Weak Symmetry Breaking (EWSB) phenomenon, no more No-Lose Theorems can be formulated and we are left, for the first time in the last 50 years, with no guarantee of new physics discoveries to be achieved in the foreseeable future. This is why the Higgs is revolutionary.

The first example of a No-Lose Theorem is the argument, schematically summarised below, that guarantees the existence of new physics beyond the Fermi theory of Weak interactions.\\
\begin{figure}[h]
\vspace{-15pt}
  \centering
  \includegraphics[width=0.8\textwidth]{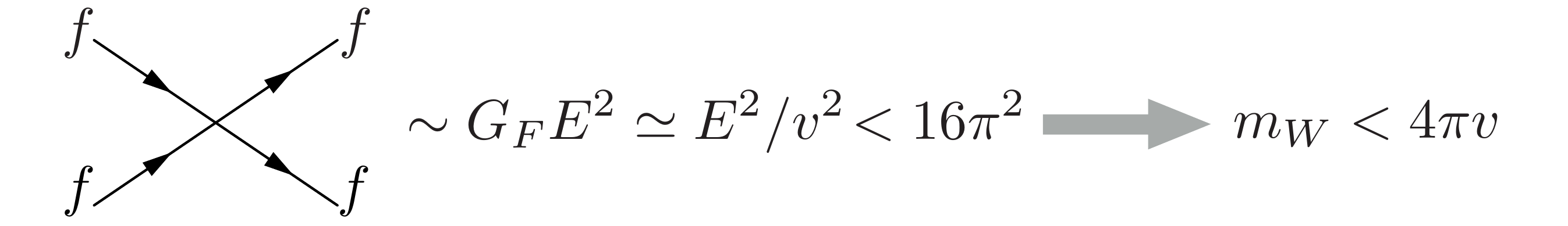} 
\vspace{-10pt}
\end{figure}

\noindent{The} point is that the Weak interactions are described, in the Fermi theory, by four-fermion operators with energy dimension equal to $6$ and thus with a coefficient (the Fermi constant $G_F$) of dimension $-2$. Dimensional analysis thus immediately revels that the $2\rightarrow2$ fermion scattering amplitudes grow, at high energies, as the square of the center-of-mass energy $E$ of the process. But the Weak scattering amplitude becoming too large, overcoming the critical value of $16\pi^2$, means that the Weak force gets too strong to be treated as a small perturbation of the free-fields dynamics and the perturbative treatment of the theory breaks down. Of course there is nothing conceptually wrong in the Weak force entering a non-perturbative regime, the problem is that this regime cannot be described by the Fermi theory, which is intrinsically defined in perturbation theory. Namely, the Fermi theory does not give trustable predictions and becomes internally inconsistent as soon as the non-perturbative regime is approached. Therefore a new theory, i.e. new physics, is absolutely needed. Either in order to modify the energy behaviour of the amplitude before it reaches the non-perturbative threshold, keeping the Weak force perturbative, or to describe the new non-perturbative regime. In all cases this new theory must show up at an energy scale below $4\pi/\sqrt{G_F}\simeq 4\pi v$, having expressed $G_F=1/\sqrt{2}v^2$ in terms of the EWSB scale $v\simeq246$~GeV. We now know that the new physics beyond the Fermi theory is the Intermediate Vector Boson (IVB) theory, which was confirmed by discovering the $W$ boson at the scale $m_W\simeq80$~GeV, far below $4\pi v$ compatibly with the Theorem.

Actually, before the $W$ discovery we already had strong indirect indications on the validity of the IVB theory and a rather precise estimate of the $W$ boson mass. However, this does not diminishes the importance of the Fermi theory No-Loose Theorem, because the latter does not rely on any hypothesis on the nature of the new physics. Namely, the Theorem guaranteed that something would have been discovered in fermion-fermion scattering, possibly not the $W$ and possibly not at a scale as low as $m_W$, even if all the theoretical speculations about the IVB theory had turned out to be radically wrong. This means in particular that if the UA$1$ and UA$2$ experiments at the CERN SPS collider had not discovered the $W$, we would have for sure continued searching for it, or for whatever new physics lies behind the Fermi theory, by the construction of higher energy machines. 

A situation like the one described above was indeed encountered in the search for the top quark, which according to a widespread belief was expected to be much lighter than $m_t\simeq 173$~GeV, where it was eventually observed. Consequently, the top discovery was expected at several lower-energy colliders, constructed before the Tevatron, which instead produced a number of negative results. However we never got discouraged and we never even considered the possibility of giving up searching for the top quark, or for some other new physics related with the bottom quark, because of a second No-Loose Theorem:\\
\begin{figure}[h]
\vspace{-15pt}
  \centering
  \includegraphics[width=0.8\textwidth]{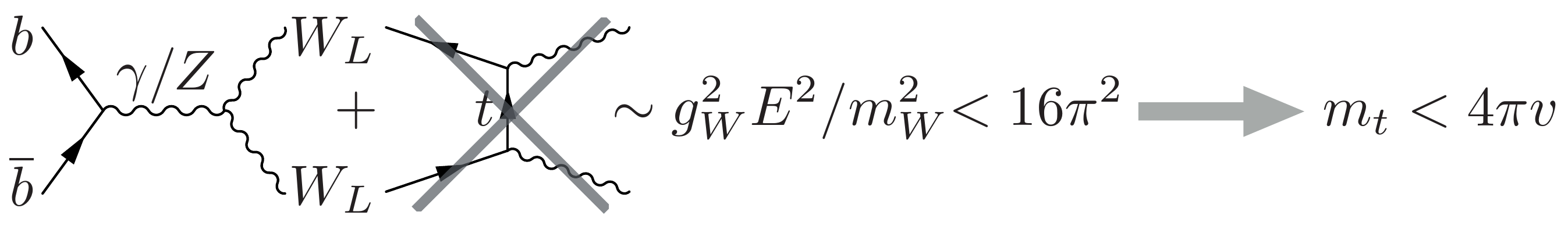} 
\vspace{-10pt}
\end{figure}

\noindent{The} Theorem relies on the validity of the IVB theory and on the existence of the bottom quark with its neutral current interactions, which we consider here as experimentally established facts at the times where the top was not yet found. The observation is that the amplitude for longitudinally polarised $W$ bosons production from a $b$ $\overline{b}$ pair grows quadratically with the energy if the top quark is absent or if it is too heavy to be relevant. It is indeed the t-channel contribution from the top exchange that makes the amplitude constant at high energies in the complete SM. Perturbativity thus requires new physics at a scale below $4\pi m_W/g_W\simeq 4\pi v$, having used the relation $m_W=g_W v/2$. When interpreted in the SM, the upper bound on the new physics scale translates in the familiar perturbativity bound on the top mass, however the Theorem does not rely on the SM and on the existence of the top quark. It states that the top, or something else, must exist beyond the bottom quark in order to moderate the growth with the energy of the scattering amplitude.

Another particle whose discovery was significantly ``delayed'' with respect to the expectations is the Higgs boson, which also comes with its own No-Loose Theorem:\\
\begin{figure}[h]
\vspace{-15pt}
  \centering
  \includegraphics[width=0.8\textwidth]{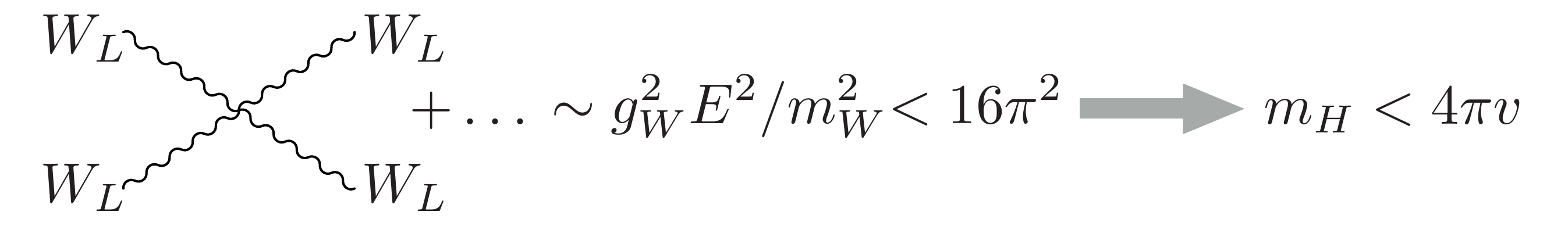} 
\vspace{-10pt}
\end{figure}

\noindent{The} growth with the energy of the longitudinally polarised $W$ bosons scattering amplitude in the IVB theory requires the presence of new particles and/or interactions, once again below the critical threshold of $4\pi v\sim3$~TeV. Given that the TeV scale is within the reach of the LHC collider, the Theorem above offered absolute guarantee of new physics discoveries at the LHC and was heavily used to motivate its construction. Now the Higgs has been found, with couplings compatible with the SM expectations, we know that it is indeed the Higgs particle the agent responsible for cancelling (at least partially, given the limited accuracy of the Higgs couplings measurements) the quadratic term in the scattering amplitude. This leaves us, as I will better explain below, with no No-Loose Theorem and thus with no guaranteed discovery to organise our future efforts in the investigation of fundamental interactions.

It is easy to understand why the SM Higgs prevents the formulation on new No-Lose Theorems by noticing that all the ones we described above are based on the growth with $E^2$ of some amplitude, a behaviour that corresponds to the presence in the theory of non-renormalizable operators with energy dimension $d=6$. The correspondence is completely transparent for the Fermi theory example, a bit less so in the other two cases where only $d=4$ interaction vertices are involved in the explicit calculations and the growth with the energy comes from the $\sim E/m_W$ behaviour of the $W_L$ polarisation vectors. In order to see the correspondence in the latter cases one first needs to uplift the theory to a formally gauge-invariant one by introducing the Goldstone boson matrix (through the Stueckelberg trick) and then to apply the Equivalence Theorem according to which the $W_L$ scattering amplitudes are equal, at high energy, to those of the corresponding Goldstone bosons. If the theory is not genuinely gauge-invariant, due to the lack of the top quark or of the Higgs, plenty of $d>4$ operators emerge after the Stueckelberg trick is applied. Some of them have $d=6$ and mediate $b{\overline{b}}\rightarrow W_L W_L$ or $W_L W_L\rightarrow W_L W_L$ producing the energy growth of the scattering amplitudes. Each time we ``exploited'' one No-Lose Theorem by discovering the corresponding particle, we got rid of one $d=6$ operator and we replaced it with one new state (e.g., the $W$ boson in the Fermi theory example) that generates the effective operator at low energies through its $d=4$ renormalizable couplings. After discovering the Higgs we are left with a genuinely gauge-invariant renormalizable theory, where no scattering amplitude can display power-like growth with the energy. No new No-Lose Theorem can thus be formulated.

The conclusion we reached at the end of the previous paragraph is essentially correct, however there are some aspects to be clarified. First of all, it is violated by gravity, because of the non-renormalizable interactions that are present in the Einstein-Hilbert action. This makes the graviton scattering amplitude grow with the energy, leading to another famous No-Loose Theorem:\\
\begin{figure}[h]
\vspace{-15pt}
  \centering
  \includegraphics[width=0.8\textwidth]{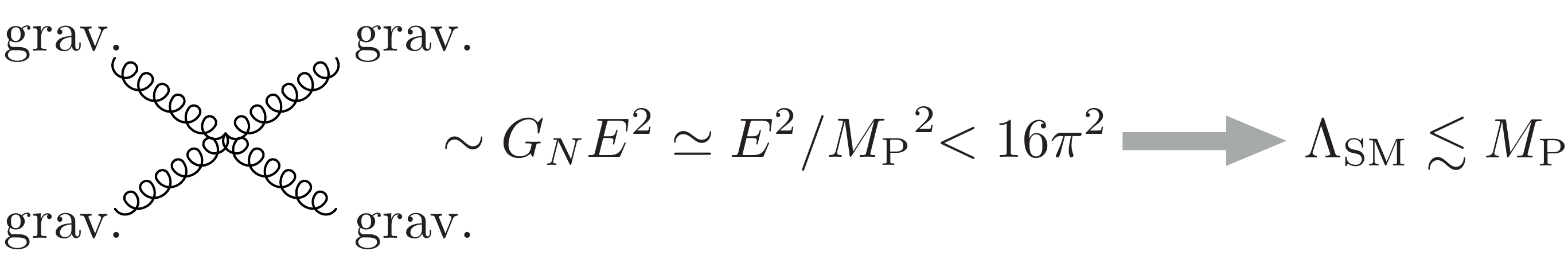} 
\vspace{-10pt}
\end{figure}

\noindent{It} predicts new physics, needed to describe quantum gravity in the high energy regime. However it leaves open the possibility that this new physics might emerge at a scale as high as the Planck mass $M_{\textrm{\tiny{P}}}\sim 10^{19}$~GeV. Given our technical inability to test such an enormous scale, it is unlikely that we might ever exploit this last No-Lose Theorem as a guide towards a concrete new physics discovery. Still, it is conceptually important as it shows that the SM is for sure not the complete theory of Nature, but rather a low-energy effective description valid below some finite energy scale $\Lambda_{\textrm{\tiny{SM}}}$ at which new particles and interactions emerge. 

The second aspect to be discussed is that even in a renormalizable theory the scattering amplitudes can actually grow with the energy. Not with a power-law, but logarithmically, through the Renormalisation Group (RG) running of the dimensionless coupling constants of the theory. The RG evolution can make some of the couplings grow with the energy until they violate the perturbativity bound, producing a new No-Lose Theorem. Obviously this No-Lose Theorem would most likely be not as powerful as those obtainable in non-renormalizable theories because the RG evolution is logarithmically slow and thus the perturbativity violation scale is exponentially high, but still it is interesting to ask if one such a Theorem exists for the SM and at which scale it points to. The answer is that perturbativity violation does not occur in the SM below the Planck mass scale, at which new physics is anyhow needed to account for gravity. Furthermore, another possibile source of inconsistency, related with the quantum stability of the vacuum, is avoided in the SM for the values we observe of its parameters. This and related aspects (in particular, the rather surprising fact that the parameters of the SM seem to be chosen on purpose such as to live at the boundary between the absolute stability and meta-stability regions) have been extensively reviewed by  F.~Bezrukov and P.~Bin\'etruy in their talks \cite{BEZRUKOV,BINETRUY} and will not be discussed here. What matters for the present considerations is that the SM can be truly extrapolated without internal inconsistencies up to $M_{\textrm{\tiny{P}}}$.

Finally, we must remember that aside from gravity there are other incontrovertible experimental facts that require the SM to be extended. New physics is needed to explain Dark Matter, neutrino masses, Inflation and Baryogenesis, however all those phenomena can be accounted for by some new physics which is too heavy or too weakly coupled to us to be directly probed in the foreseeable future. Namely, those evidences of new physics produce no new No-Loose Theorem and thus no guarantees of future discoveries.

\subsection*{The Naturalness Argument}

While far from offering any guarantee of a concrete new physics discovery, the Naturalness Argument is the closest thing we have to a No-Loose Theorem. It can be formulated as follows. By the Higgs discovery we got rid of all $d>4$ operators in our theory, but we introduced one new operator of dimension $d<4$: the Higgs mass-term
\begin{equation}
\mu^2 H^\dagger H\,,
\end{equation}
where $\mu=m_H/\sqrt{2}\simeq 88$~GeV for $m_H\simeq125$~GeV. Operators with $d<4$ have $d>0$ coefficients that generically suffer, in the absence of a symmetry protection, of a Naturalness problem. Namely, it is hard to explain how their low value can be generated by a microscopic UV dynamics characterised by a much higher typical scale. The Higgs mass-term $\mu^2$ is an input parameter of the SM, therefore the microscopic explanation of its origin must be attributed to new physics at the SM cutoff $\Lambda_{\textrm{\tiny{SM}}}$ that we saw above to be potentially as high as the Planck scale. But it was so, what could give origin to this huge hierarchy of scales? Namely
\begin{equation}
{\textrm{why}}\;\;\frac{\mu^2}{\Lambda_{\textrm{\tiny{SM}}}^2}\lll1\;{\textrm{\Large{?}}}
\end{equation}
Not to suffer from this problem, the new physics should be much lighter than $M_{\textrm{\tiny{P}}}$, possibly as light as the TeV scale and thus directly testable at the LHC.

There is a couple of misconceptions about Naturalness that is worth discussing. First of all, Naturalness is not an internal inconsistency or a problem for the pure SM theory because the Higgs mass-term is merely an input parameter of the SM, to be fixed by observations. Therefore it makes no sense to discuss about its value, which is not predicted, diverges and is ultimately set by a renormalisation condition. The problem can only be formulated if we worry about the microscopic origin of the Higgs mass-term, namely if we postulate the existence of a more fundamental theory where the Higgs mass-term emerges as a low-energy effective operator, with a coefficient predicted by the fundamental theory in terms of its own more fundamental input parameters. This theory might not exist, but this would mean that the Higgs mass-term (and in turn the EWSB scale) does not have a microscopic origin and thus it will forever remain with us, at all scales, as a fundamental constant of Nature. While admittedly implausible, we cannot discard this possibility. Second, the importance of the Naturalness Problem is not at all diminished by the SM Higgs discovery. It is the exact contrary because it is precisely the existence of the Higgs what poses the Naturalness Problem. We learned that the Problem is real at the same moment when we discovered that the Higgs is real.

The Naturalness (or Hierarchy) Problem is more than exhaustively discussed in the literature. The reader is referred to two recent essays \cite{Giudice:2008bi,Barbieri:2013vca} for a deeper discussion, references and various alternative formulations. Here I will merely recall the familiar semi-quantitative formulation in terms of the level of fine-tuning $\Delta$ needed to account for the observed value of the Higgs mass from new physics at the cutoff scale $\Lambda_{\textrm{\tiny{SM}}}$. The idea is that in the fundamental theory formula that predicts the Higgs mass-term (or, which is the same, the squared Higgs mass $m_H^2=2\mu^2$) it will be possible to identify two contributions. One of them, $\delta_{\textrm{\tiny{SM}}} m_H^2$, results from the exchange of virtual quanta with virtuality below  $\Lambda_{\textrm{\tiny{SM}}}$. The other one, $\delta_{\textrm{\tiny{BSM}}} m_H^2$, comes from physics above $\Lambda_{\textrm{\tiny{SM}}}$. Namely, we have
\begin{equation}
m_H^2=\delta_{\textrm{\tiny{SM}}} m_H^2+\delta_{\textrm{\tiny{BSM}}} m_H^2\,.
\end{equation}
The BSM contribution is unpredictable as long as we don't specify the more fundamental BSM theory. The SM one can instead be estimated because the BSM theory reduces by assumption to the SM below the cutoff. The SM particles, and in particular the top quark with its sizeable Yukawa coupling with the Higgs, $y_t\sim1$, produces a contribution of order 
\begin{equation}
\delta_{\textrm{\tiny{SM}}} m_H^2\simeq \frac{3\,y_t^2}{8\pi^2}\Lambda_{\textrm{\tiny{SM}}}^2\,,
\end{equation}
which is much larger than the observed $m_H^2$ if $\Lambda_{\textrm{\tiny{SM}}}$ is significantly above the TeV. The BSM term $\delta_{\textrm{\tiny{BSM}}} m_H^2$ is thus obliged to be almost equal and with opposite sign in order to get $m_H^2$ right. This results in a cancellation, or fine-tuning, or order 
\begin{equation}
\label{tuning}
\Delta \simeq \frac{\delta_{\textrm{\tiny{SM}}} m_H^2}{m_H^2}\simeq \frac{3\,y_t^2}{8\pi^2}\Lambda_{\textrm{\tiny{SM}}}^2\simeq\left(\frac{\Lambda_{\textrm{\tiny{SM}}}}{450\,{\textrm{GeV}}}\right)^2\,,
\end{equation}
where the observed value of the Higgs mass, $m_H\simeq125$~GeV, has been used. 

We call ``Natural'' the situation where $\Delta\lesssim1$, ``Un-Natural'' the one in which $\Delta\gg1$. It is impossible to give the sharp upper bound on $\Delta$ above which the theory qualifies as Un-Natural. Anybody can set his/her own personal threshold provided she/he understands what a given level of tuning concretely means and how it affects the actual predictability of the Higgs mass. If $m_H^2$ results from the cancellation of two unrelated terms which are both a factor of $\Delta$ larger than $m_H^2$ in absolute value, achieving even just an order one estimate of the Higgs mass requires an accuracy of order $1/\Delta$ in the determination of the two terms. It might be impossible to achieve this accuracy even if we would eventually came to know all the details of the fundamental theory. If $\Lambda_{\textrm{\tiny{SM}}}=M_{\textrm{\tiny{SM}}}$, so that $\Delta\sim 10^{32}$, a $32$-digit cancellation is taking place between $\delta_{\textrm{\tiny{SM}}} m_H^2$ and $\delta_{\textrm{\tiny{BSM}}} m_H^2$ and the actual prediction for $m_H$ emerges from the $33$rd digit. If the cancellation is too large to be overcame, which is probably the case already for $\Delta=100$, we will face the practical impossibility to predict $m_H$ even if the fundamental theory formally allows for it. The Higgs mass would thus remain an input parameter and no concrete progress would be made on its fundamental origin.

This makes extremely important to search at the LHC for new physics associated with the microscopic origin of $m_H$ and of the EWSB scale. If it is at the TeV, so that the tuning $\Delta$ in Eq.~(\ref{tuning}) is not much larger than $1$, we have good chances to discover it. If nothing is found, strong lower limits will be set on $\Lambda_{\textrm{\tiny{SM}}}$ and consequently on the level of tuning. In this case, Un-Naturalness will be discovered and we will be forced to reconsider the problem of the EWSB scale generation. Know options are giving to the EWSB scale an environmental origin through Landscape and Vacua Selection, or explaining it ``dynamically'' through the cosmological evolution as in the recent ``Relaxion'' proposal \cite{Graham:2015cka}. The very existence of those interesting speculative ideas, clearly summarised by D.~E.~Kaplan in his talk at this conference \cite{KAPLAN} (see also \cite{BINETRUY}), gives us the feeling of how radically the discovery of Un-Naturalness would change our perspective on the physics of fundamental interactions. Searching for Naturalness at the LHC is thus, after all, a No-Lose game because it will give us extremely valuable informations even if nothing will be found.

\section{Higgs Physics and BSM}\label{s2}

The discovery of the Higgs boson and the verification of its SM-like properties, whose far-reaching generic implications I described above, is the main achievement of the experimental exploration of the scalar Higgs sector at the LHC (see the talk by P.~Savard \cite{SAVARD} for a comprehensive review). However valuable results were also obtained in the study of specific BSM scenarios through the measurement of Higgs couplings and the search for extra scalars. The impact of Higgs sector results on concrete BSM ideas is summarised below.

\subsection*{The Composite Higgs}

The composite Higgs scenario \cite{Panico:2015jxa} is the idea that the Higgs boson could be the bound state of a new strongly-interacting dynamics characterised by a confinement scale $m_*\sim\;$TeV. If this was the case, the whole Higgs Lagrangian and in particular the mass-term would originate from physics at the scale $m_*$. It would receive no contributions from much higher scales and the Naturalness Problem would be avoided. The confinement scale $m_*$ is expected to be itself Natural as it is set, like it happens for the QCD scale $\Lambda_{\textrm{\tiny{QCD}}}$, by the mechanism of Dimensional Transmutation.

Measuring the Higgs boson couplings and searching for deviations from the predictions of the SM, in which the Higgs is an elementary point-like particle, is, rather obviously, a way to test its possible composite nature. After all, it is by observing modifications of its coupling to the photon that we discovered that the proton is a composite object. Much less obvious is instead that quite sharp theoretical predictions can be made for the pattern of Higgs coupling modifications that are expected in the composite Higgs scenario, in spite of the fact that a strongly-interacting confining dynamics is involved. This is because the composite Higgs cannot be a ``generic'' bound state of the composite sector, but a ``specific'' one: a pseudo-Nambu--Goldstone Boson (pNGB) associated with the spontaneous breaking of a global symmetry of the composite sector. This need mainly comes from the observation that otherwise we would not understand why \mbox{$m_H\ll m_*\sim\;$TeV}, while this scale separation is completely Natural for a pNGB Higgs, whose mass is protected by the Goldstone symmetry. The pNGB nature of the Higgs and the Goldstone symmetry (plus additional assumptions specified below) lead to predictions for the Higgs couplings to vector bosons and fermions in terms of a single parameter 
\begin{equation}\label{xi}
\xi=\frac{v^2}{f^2}\,.
\end{equation}
We see that $\xi$ is defined in terms of the ratio between the EWSB scale \textrm{$v\simeq246\;$GeV} and a newly introduced dimensionful  parameter $f$ that represents the decay constant of the pNGB Higgs. The equivalent of $f$ in an analogy with the QCD pions would be the pion decay constant $f_\pi$.

The modification of the Higgs couplings to $W$ and $Z$ vector bosons with respect to the SM prediction, expressed in terms of the habitual $\kappa$ factor, is given by
\begin{equation}
\kappa_V=\frac{g_{hVV}^{\textrm{\tiny{CH}}}}{g_{hVV}^{\textrm{\tiny{SM}}}}=\sqrt{1-\xi}\,.
\end{equation}
Notice that $\xi$ ranges from $0$ and $1$, so that the expression above never assumes imaginary values and is always smaller than one. The modified Higgs couplings to fermions are less sharply predicted because they are subject to additional discrete model-building ambiguities related with the choice of the quantum numbers of the composite sector operators that couple with the SM fermions. The most commonly adopted choices define two scenarios, denoted as \mbox{MCHM$_{4}$} and \mbox{MCHM$_{5}$}, and produce predictions 
\begin{eqnarray}
&&\kappa_F^{\mathbf{4}}=\sqrt{1-\xi}\,,\\
&&\displaystyle\kappa_F^{\mathbf{5}}=\frac{1-2\xi}{\sqrt{1-\xi}}\,,\nonumber
\end{eqnarray}
but few other options might be also considered.\footnote{For instance, it is easy to find models where $\kappa_u\neq\kappa_d$. Or models where extra Higgs scalars are present and contribute, through mixing, to the coupling deviations. This latter option should be studied in connection with direct searches of extra scalar, a subject that is not much developed in the composite Higgs framework. Finally, it should be kept in mind that the $\kappa$'s are not necessarily universal for all the quark families and for the leptons. This is immaterial for the time being because third family quarks $\kappa_{t,b}$ are what drives the current fit.}
 Intuitively, the reason why the Higgs coupling to $W$ and $Z$ is uniquely predicted while the one to fermions is not is that the EW bosons are unavoidably introduced in the theory as gauge fields and therefore they necessarily couple to the composite sector through its global current operators associated with the SM symmetry group. The quantum numbers of the current are uniquely fixed, while the quantum numbers of the composite sector operators the fermions couple to are not fixed and different viable options exist. 

\begin{figure}[t]
\centering
\includegraphics[width=.6\textwidth]{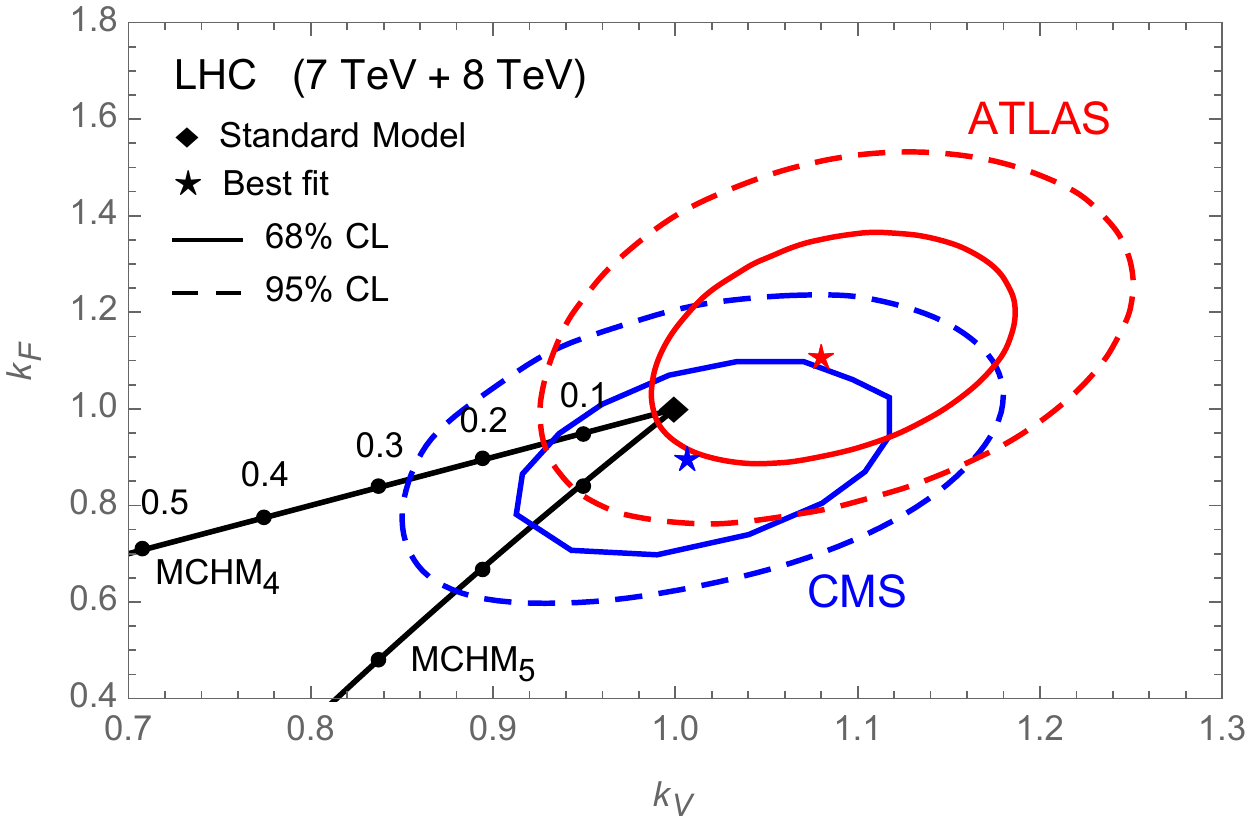}
\caption{Fit of the Higgs coupling strength to the gauge bosons ($k_V$) and fermions ($k_F$) obtained by the
ATLAS (red contours) and CMS collaborations (blue contours) from the combination of the $7$ and $8\ \mathrm{TeV}$ LHC data.
The solid black lines show the predictions in the \mbox{MCHM$_{5,4}$} models for different values of $\xi$.}
\label{fig:kvkf}
\end{figure}

Higgs coupling modifications in the $\kappa_V$--$\kappa_F$ plane have been searched for by both ATLAS \cite{Aad:2015gba} and CMS \cite{Khachatryan:2014jba}, with the results displayed in Fig.~\ref{fig:kvkf} and compared with the composite Higgs predictions as a function of $\xi$. The contours on the figure are not exclusion lines, therefore the figure cannot be directly used to infer the upper limit on $\xi$ and a dedicated statistical analysis is needed. This was performed by ATLAS in Ref.~\cite{Aad:2015pla}, obtaining $\xi<0.12$ in the  \mbox{MCHM$_{4}$} and $\xi<0.10$ in the  \mbox{MCHM$_{4}$} at $95\%$ CL. The limit is significantly stronger than expected and stronger than the one obtainable from the CMS results because the central value of the ATLAS measurement sits at $\kappa_V$ and $\kappa_F$ larger than one, opposite to the composite Higgs expectations. This bound on $\xi$ singles out as the stronger constraint we currently have on the composite Higgs scenario. Indeed, it is true that even stronger bounds on the Higgs couplings come from their indirect radiative effects on ElectroWeak Precision Tests (EWPT) observables, but it is also true that coupling modifications are not the only sources of corrections to the EWPT in the composite Higgs scenario, nor the dominant ones. This makes that the EWPT limits on Higgs couplings are rather easily evaded by concrete composite Higgs models, while the LHC ones are model-independent and unavoidable.

On top of this, coupling modifications are of outmost importance in the composite Higgs scenario because they are directly connected with the amount of fine-tuning (i.e., the degree of Un-Naturalness) of the theory, which is definitely the most important quantity to be kept under control in a construction that aims to address the Naturalness Problem. The coupling/tuning connection comes from the fact that coupling deviations are sensitive to the parameter $\xi$, which sets a bound
\begin{equation}\label{tunxi}
\displaystyle
\Delta\gtrsim\frac1{\xi}\,,
\end{equation}
on the level of tuning. Intuitively, this bound on $\Delta$ emerges because $\xi$ (\ref{xi}) measures the ratio between two symmetry breaking scales $v$ and $f$, respectively associated with the breaking of the EW symmetry and of the Goldstone symmetry from which the Higgs emerges. One would expect the two scales to be similar, while a certain amount of ``Un-Natural'' hierarchy is required to make $\xi$ small. More technically, the problem is that the Higgs field always enters the scalar potential in the combination $H/f$, so that its minimum generically sits at $v\sim f$ producing $\xi\sim1$. Reducing the Higgs VEV requires reducing the mass-term relative to the quartic by a factor of $\xi$, giving origin to the estimate in Eq.~(\ref{tunxi}). Notice that the estimate only takes into account a single source of fine-tuning, which is generically present in all models, related with the size of the mass-term relative to the quartic. An extra amount of Un-Natural cancellations can be needed in specific cases to produce a realistic Higgs VEV and mass. This is why Eq.~(\ref{tunxi}) only provides a lower bound on the total amount of fine-tuning.

\begin{figure*}[t!]
\begin{center}
\includegraphics[scale=0.3]{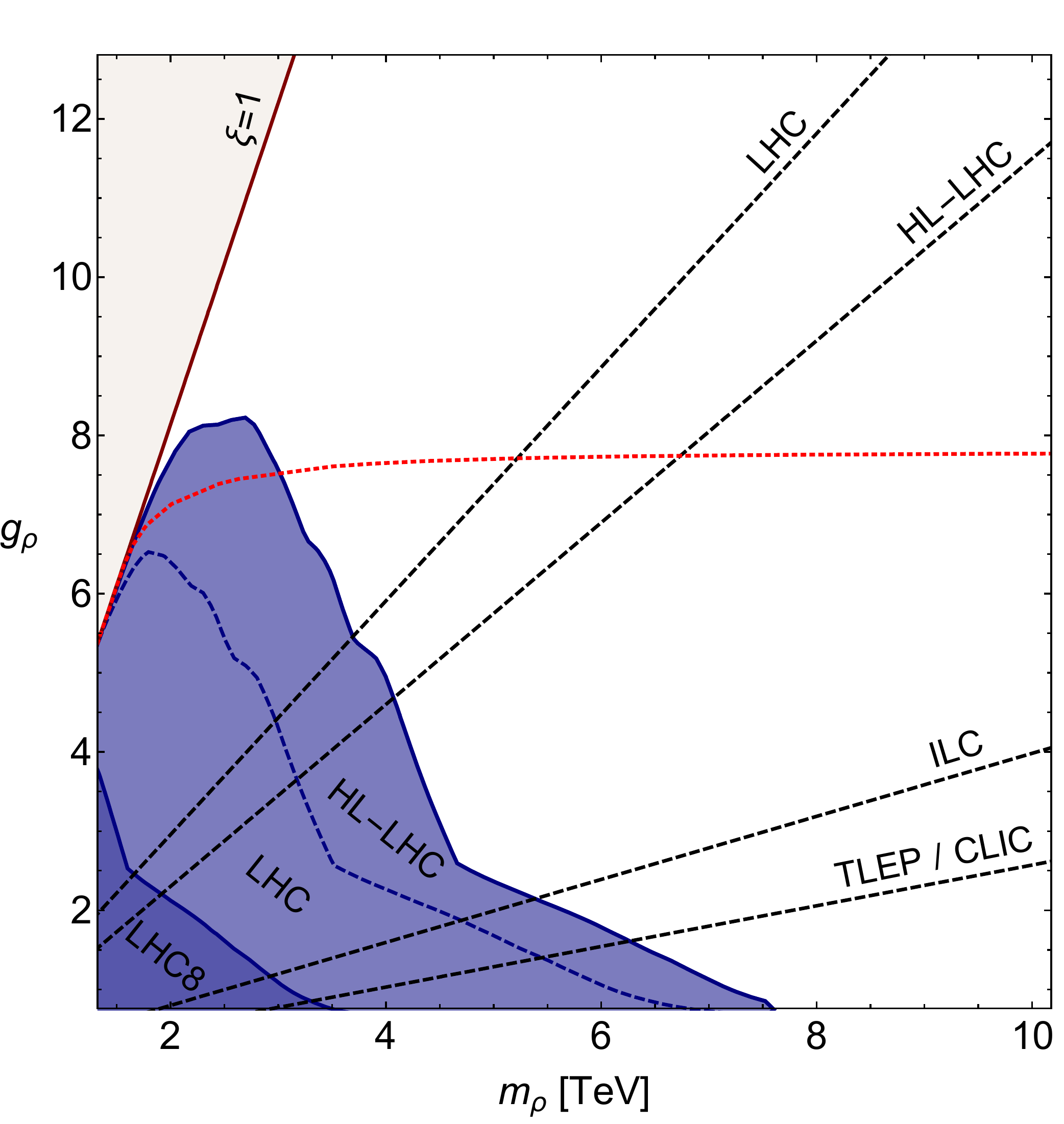}\hspace{4mm}
\hspace{0.0cm}\includegraphics[scale=0.3]{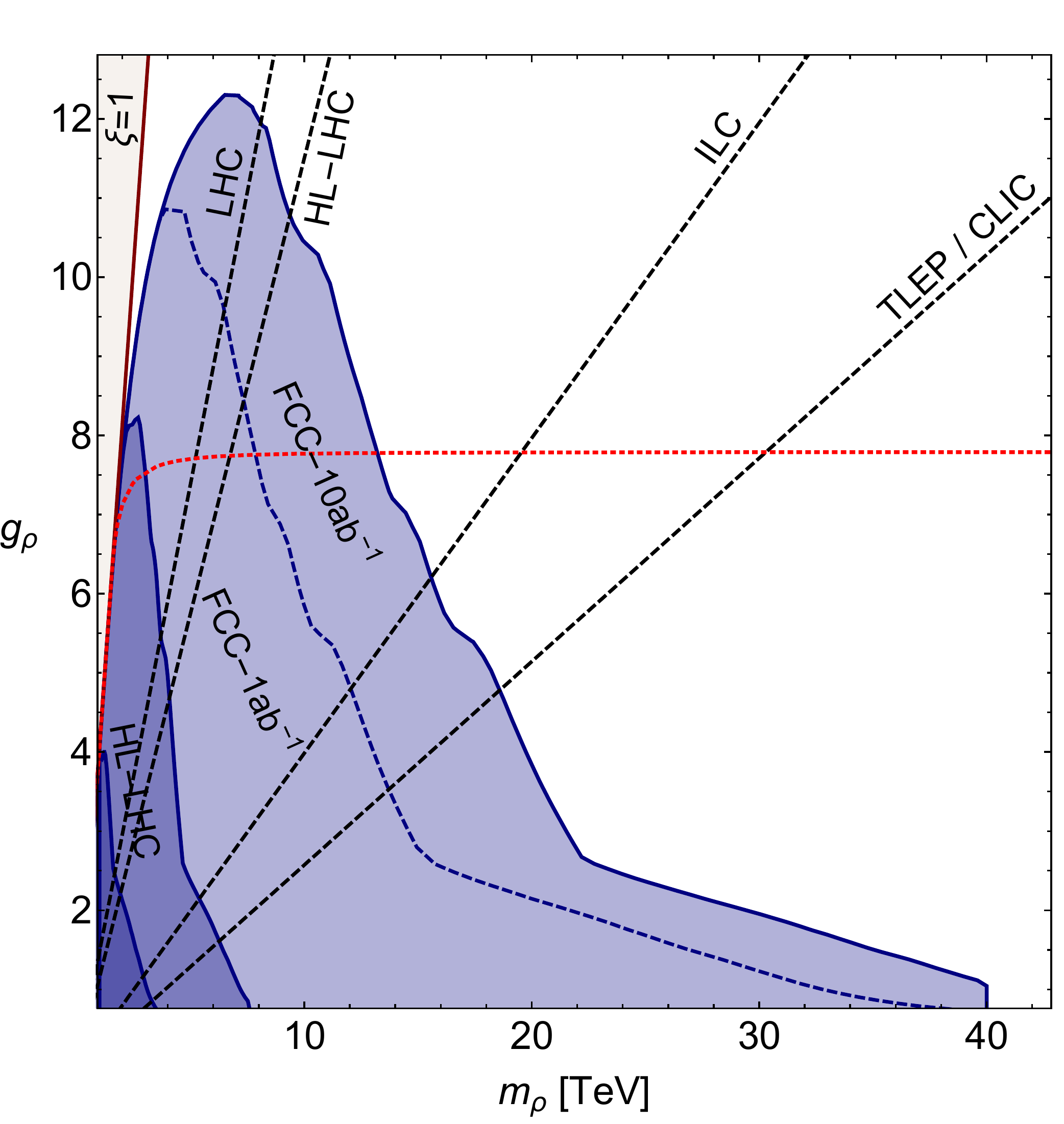}
\caption{\small Comparison of direct and indirect searches for several existing and future colliders, among which the FCC-hh option at $100$~TeV, from Ref.~\cite{Thamm:2015zwa}. The regions above the dashed lines are excluded by coupling measurements performed at different colliders, the blue contours are the direct search constraints.
}\label{fig:mrho_grho_plane}\vspace{-0.5cm}
\end{center}
\end{figure*}

The expected accuracy of Higgs couplings determination at the LHC and at future colliders and the corresponding reach in $\xi$ have been extensively studied by several groups. The result (see e.g. \cite{Thamm:2015zwa} for a summary) is that an exclusion reach $\xi\lesssim0.1$ is expected at the $13$-TeV LHC with $300\;{\textrm{fb}}^{-1}$. A mild improvement will come from the high-luminosity (HL-LHC) program while new machines such as the ILC, TLEP or CLIC are needed to extend the reach by one order of magnitude. Given that the observed ATLAS bound is already at the level of $\xi\lesssim0.1$, no much progress in the limit is expected at the LHC or at the HL-LHC. Furthermore, achieving a discovery appears impossible. On the contrary, the sensitivity to the composite Higgs scenario through the direct search for new particles will rapidly increase with the energy and the luminosity. A comparative study of the direct and indirect exclusion reach from, respectively, resonance searches and coupling measurements has been performed in Ref.~\cite{Thamm:2015zwa}, with the result displayed in Fig.~\ref{fig:mrho_grho_plane}. The plot refers to the search for spin-one resonances ({\it{i.e.}}, other bound states of the composite sector from which the Higgs emerges) in the triplet of the EW group, whose phenomenology is characterised, aside from their mass $m_\rho$, by a  coupling $g_\rho$ that measures the strength of the composite sector interactions. The form of the exclusion contours can be understood (see \cite{Thamm:2015zwa} and references therein for details) by the fact that the coupling of the resonance with the Higgs and the vectors bosons is proportional to $g_\rho$, while the one to light quarks and leptons is inversely proportional to $g_\rho$. In particular, this suppresses the production rate at large $g_\rho$ and makes the limit disappear. The composite sector coupling strength can be estimated as $g_\rho=m_\rho/f$, from which we obtain the relation, used for the comparison, with the limit on $\xi=v^2/f^2$ from coupling measurements.

\subsection*{The SUSY Higgs}

\begin{figure*}[t!]
\begin{center}
\includegraphics[width=200pt]{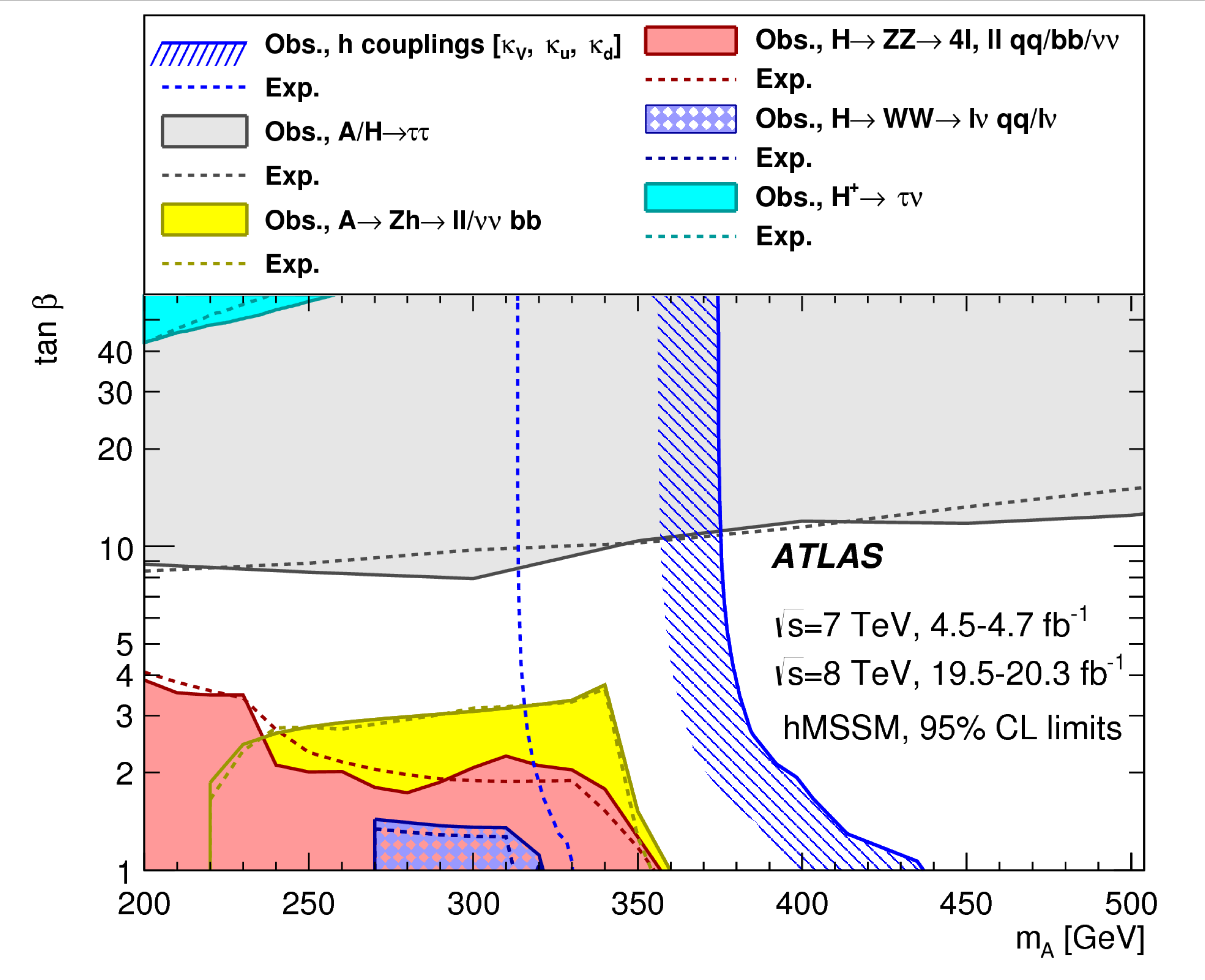}\hspace{4mm}
\hspace{0.0cm}\includegraphics[width=200pt]{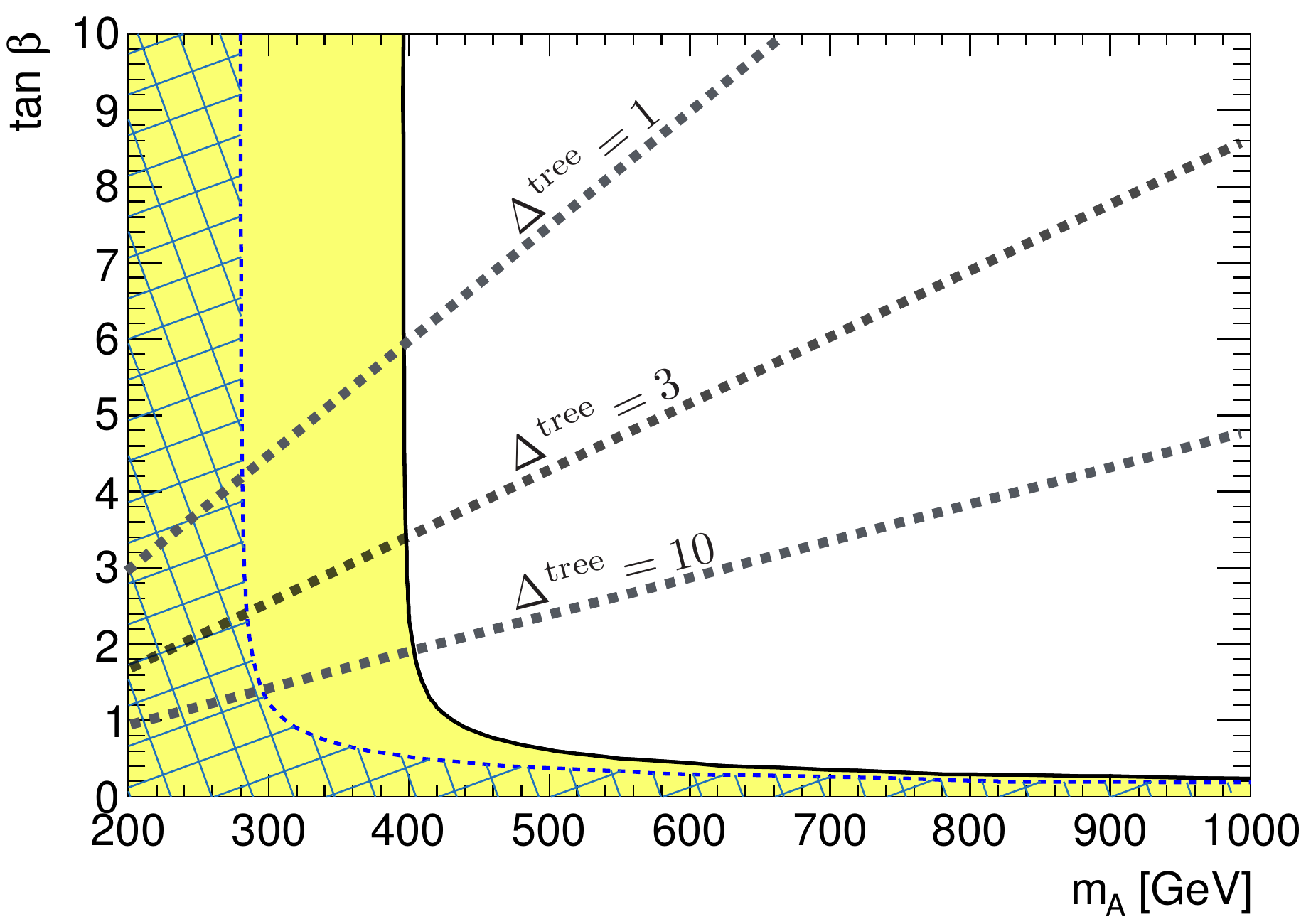}
\caption{\small Left panel: the constraints on the MSSM from Higgs physics as reported in Ref.~\cite{Aad:2015pla}. Right panel: bounds from Higgs coupling measurements (from a plot in a preliminary version of Ref.~\cite{Aad:2015pla}) on which I overlaid the ``tree-level'' tuning contours.}\label{fig:MSSM}\vspace{-0.5cm}
\end{center}
\end{figure*}

Low-energy supersymmetry (SUSY) is the most popular framework to address the Naturalness Problem and it can also be studied through the experimental exploration of the scalar sector. Indeed, SUSY requires the Higgs sector to be extended with respect to the SM, containing at least two Higgs doublets in order to give mass to both the up- and down-type quarks. This produces extra scalar bosons, to be directly searched for, and modifications of the $125$~GeV Higgs couplings. In the Minimal Supersymmetric SM (MSSM), or more precisely in its simplified version defined in Ref~\cite{Aad:2015pla} where one term is neglected in the scalar mass-matrix, the $\kappa$'s read
\begin{eqnarray}
&\displaystyle
\kappa_u=\frac{\cos\alpha}{\sin\beta}\simeq 1 -\frac1{1+t_\beta^2}\varepsilon\,,\;\;\;\;\;\kappa_u=-\frac{\sin\alpha}{\cos\beta}\simeq 1 +\frac{t_\beta^2}{1+t_\beta^2}\varepsilon\,,\;\;\;\;\;
 \kappa_V=\sin{(\beta-\alpha)}\simeq 1-{\mathcal{O}}(\varepsilon^2)\,, &\nonumber \\
&\displaystyle \tan\alpha=\frac{(m_A^2+m_Z^2)t_\beta}{m_h^2(1+t_\beta^2)-m_Z^2-m_A^2 t_\beta^2}\simeq-\frac1{t_\beta}+{\mathcal{O}}(\varepsilon)\,,&\label{hcMSSM}
\end{eqnarray}
where $t_\beta=\tan\beta$ is the ratio between up- and down-type Higgs VEVs, $m_A$ is the mass of the CP-odd extra Higgs scalar and $m_h=125$~GeV. The expansion for $\varepsilon=m_h^2/m_A^2\ll1$ is also reported in the equation. This formula allows us to set limits from the Higgs coupling measurements in the $(m_A,\tan\beta)$ plane, on which the bounds from direct searches of extra scalars can also be reported. The result, obtained by ATLAS in Ref.~\cite{Aad:2015pla}, is shown in the left panel of Fig.~\ref{fig:MSSM}.

The bounds disappear for large $m_A$ because the extra scalars become too heavy to be produced and the couplings (\ref{hcMSSM}) approach their SM values. This behaviour is of course not surprising, and it is common to (almost) all BSM scenarios. Namely, one can (almost) always decouple the new physics and reproduce the SM as accurately as needed by the data. However, this can be normally achieved only at the price of some amount of Un-Natural fine-tuning, like we saw in the previous section for the composite Higgs scenario. Instead, decoupling the extra scalars is technically Natural in the MSSM, in the following sense. Reproducing the correct EWSB scale from the minimisation of the potential requires imposing on the parameters of the theory a certain condition, which for $t_\beta\gg1$ reads \footnote{In the equation that follows, $\mu$ is the SUSY mass, $m_u$ is the soft up-type mass and $\delta$ accounts for the radiative contribution to the Higgs quartic in the notation of Ref.~\cite{Aad:2015pla}.}
\begin{equation}\label{minMSSM}
\displaystyle \frac{m_Z^2}2\simeq\frac{m_A^2}{t_\beta^2}-{\widetilde{m}}_u^2\,,\;\;\;\;\;{\textrm{where}}\;\;\;{\widetilde{m}}_u^2=\mu^2+m_u^2+\delta\,.
\end{equation}
As $m_A$ increases, the first term in the $Z$ mass formula becomes larger and larger and a more and more accurate cancellation has to be enforced with the second term in order to reproduce $m_Z$. If however $\tan\beta$ also increases, such as to keep $m_A/t_\beta$ at the EW scale, the first term remains small and no fine-tuning is required. More technically, the level of tuning can be estimated as
\begin{equation}\label{Dtree}
\displaystyle
\Delta^{\textrm{tree}}=\frac{(m_A/t_\beta)^2}{(m_Z/\sqrt{2})^2}\simeq\left(\frac{6}{t_\beta}\right)^2 \left(\frac{m_A}{400\,{\textrm{GeV}}}\right)^2 \,,
\end{equation}
where $m_A=400$~GeV, which is representative of the current limit, has been chosen as reference. Contour lines of $\Delta^{\textrm{tree}}$ in the $(m_A,\tan\beta)$ plane are shown on the right panel of Fig.~\ref{fig:MSSM} and compared with the experimental limits. We see that low-tuning configurations exist even for very large $m_A$, where no bound can be set neither from the coupling measurements nor from the direct searches. Therefore in the MSSM, differently from the case of the composite Higgs, there is no direct connection between new physics in the scalar sector and the level of Un-Naturalness of the theory. Admittedly, this makes scalar sector physics slightly less interesting in the MSSM.

However the MSSM, as we now know for sure after the direct measurement of $m_h\simeq125$~GeV, is not the appropriate model to discuss Naturalness in the SUSY context because a much larger source of fine-tuning exists than the ``tree--level'' one we accounted for by Eq.~(\ref{Dtree}). The well-known problem is that $m_h$ is smaller than $m_Z$ at tree-level and making it large enough requires a sizeable radiative correction to the Higgs quartic term in the potential. This correction grows logarithmically with the mass of the stops and is unavoidably accompanied by a correction to the up-type mass-term $m_u^2$ that instead grows like the stop mass squared. This needs to be canceled not to produce an unacceptably large second term in Eq.~(\ref{minMSSM}), resulting in a large fine-tuning. The estimate of Ref.~\cite{Hall:2011aa} is that stops as heavy as at least $1$~TeV are needed and the tuning is $\Delta>100$. Therefore the MSSM is not a Natural theory and furthermore we have little chances to discover it at the LHC given that the entire SUSY spectrum could be above the TeV.\footnote{For a fair comparison with the composite Higgs scenario, I discussed in the talk (but I don't have space to report on this here) how the $125$~GeV Higgs discovery also had an impact on composite Higgs constructions, though not as dramatic as on the MSSM. Namely, the relatively light Higgs mass obliges certain particles, the ``Top Partners'', to be light and within the LHC reach \cite{Matsedonskyi:2012ym}. This is another important BSM implication of Higgs physics.}

\begin{figure*}[t!]
\begin{center}
\includegraphics[width=180pt]{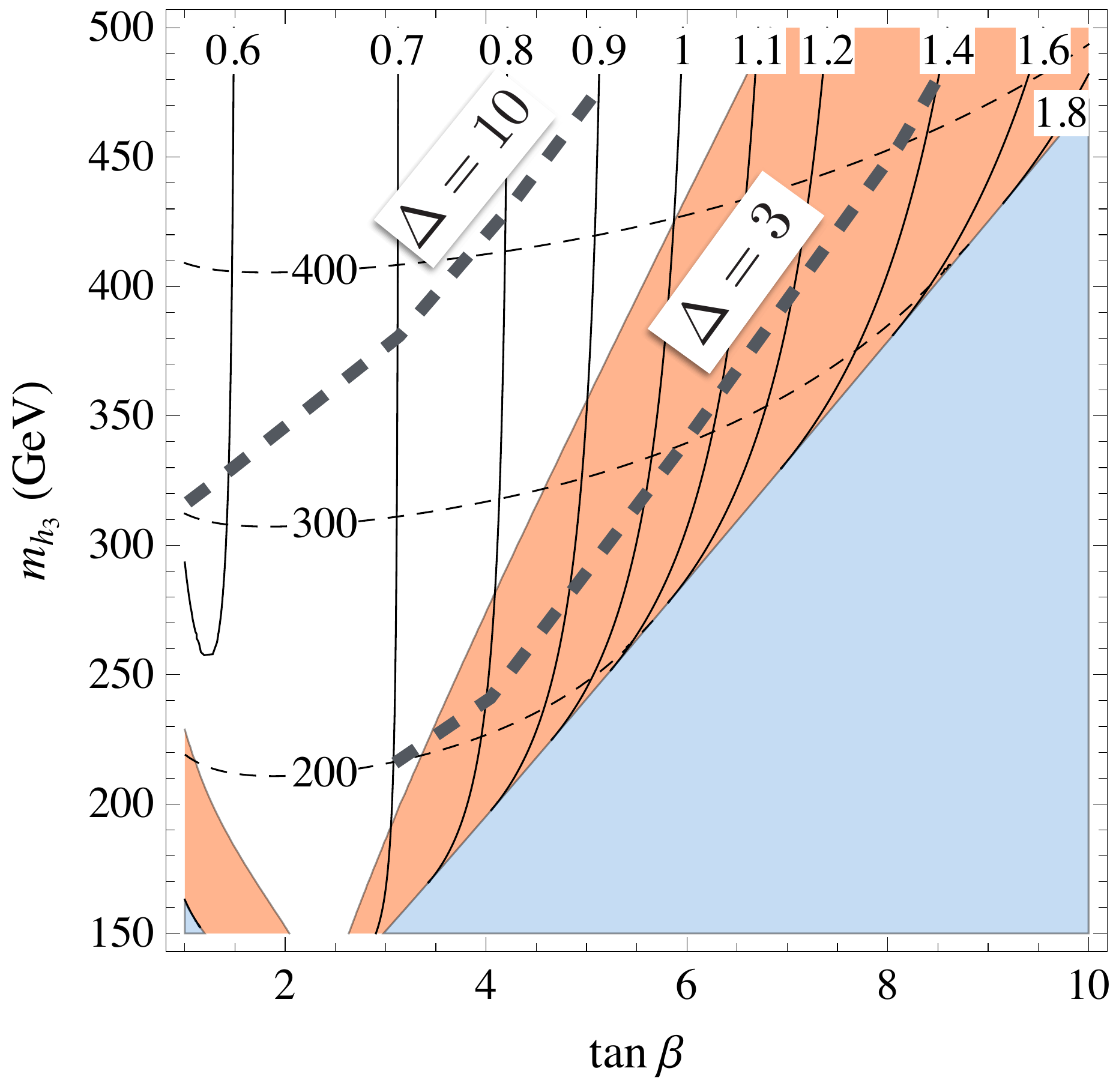}\hspace{4mm}
\hspace{0.0cm}\includegraphics[width=180pt]{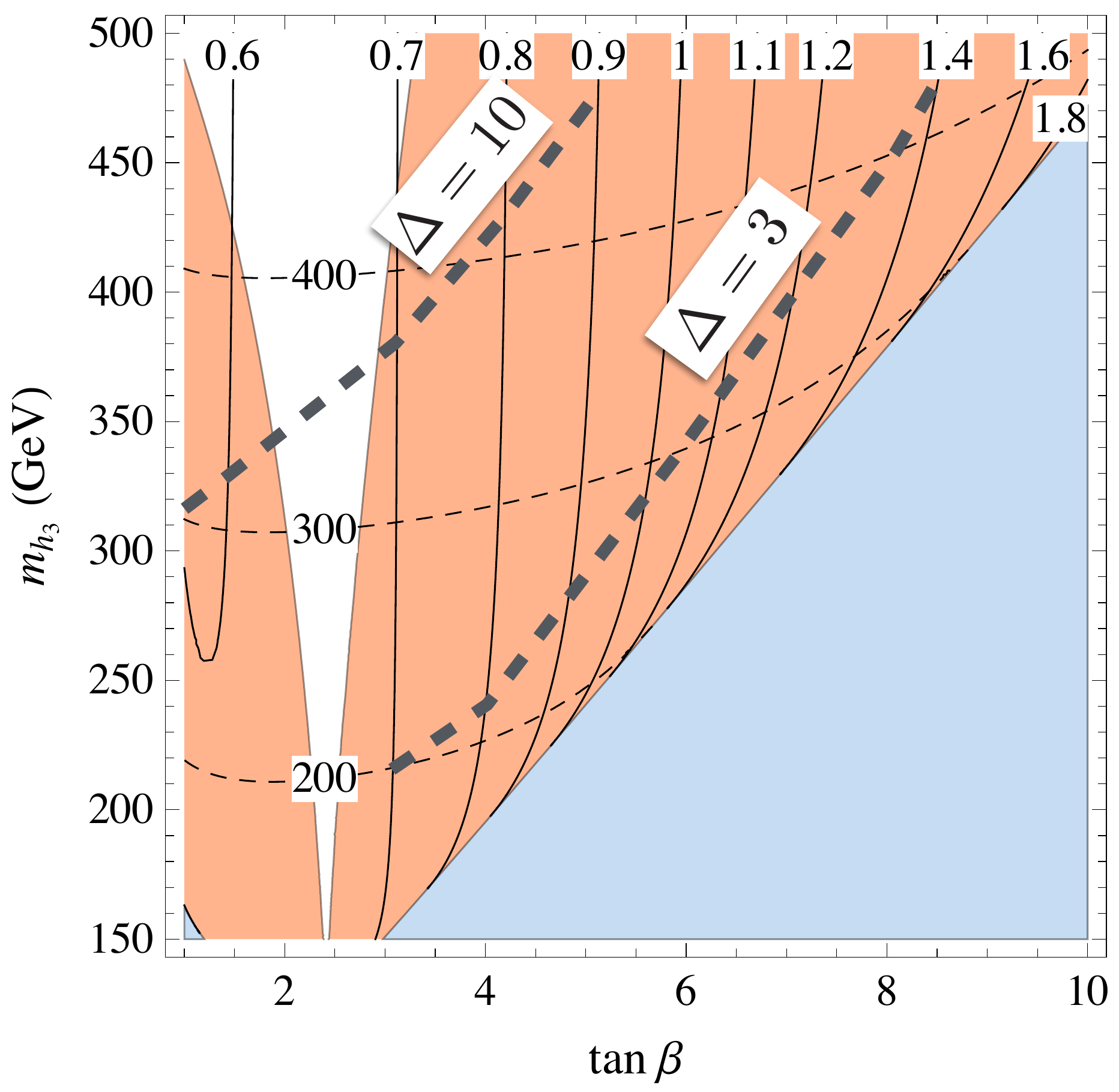}
\caption{\small Bounds \cite{Barbieri:2013nka} on \mbox{$\lambda$SUSY} from Higgs coupling measurements in the plane $(\tan\beta,m_{h_3})$ where $m_{h_3}=m_H$ is the mass of the CP-even Higgs. The plots refers to the limiting situation where the additional CP-even scalar from the extra singlet is decoupled. Thin continuous and dashed lines denote $m_{H^\pm}$ and $\lambda$ contours, respectively. The left panel shows current limits from the $7$ and $8$~TeV results of the LHC, the right panel is a projection of the $13$~TeV reach.}\label{fig:lSUSY}\vspace{-0.5cm}
\end{center}
\end{figure*}

The considerations above strongly motivate the study of alternative SUSY models, among which the \mbox{$\lambda$SUSY} framework \cite{Barbieri:2006dq} emerges as a particularly plausible option. In \mbox{$\lambda$SUSY}, an extra singlet chiral super-multiplet $S$ is added to the MSSM and coupled through a term $\lambda S H_u H_d$ in the superpotential. This gives a new contribution to the Higgs quartic term that can produce, if $\lambda\gtrsim1$, a large enough Higgs mass already at the tree-level, with no need of heavy stops and large fine-tuning from radiative corrections. To be precise, and this is very important for the considerations that follow, the mechanism requires not only sizeable $\lambda$, but also moderate $\tan\beta$ below around $10$. Therefore the Natural decoupling limit of the MSSM, with large $\tan\beta$, cannot be taken in \mbox{$\lambda$SUSY} and a more direct connection is expected between Naturalness and Higgs physics. This is quantitatively illustrated by the Higgs VEV formula, which in the rough approximation of neglecting the gauge contribution to the potential reads
\begin{equation}
\displaystyle
\frac{\lambda^2v^2}2=m_A^2-m_{H^\pm}^2\,.
\end{equation}
We see that now, differently from the equivalent MSSM expression (\ref{minMSSM}), the masses of the pseudo-scalar and charged Higgses enter with no suppression factor and the tuning estimate reads \footnote{I will need to express $\Delta$ in terms of $m_{H^\pm}$ rather than $m_A$ in order to be able to draw the tuning contours in Fig.~\ref{fig:lSUSY}.}
\begin{equation}
\label{Dl}
\displaystyle
\Delta=\frac{(m_{H^\pm})^2}{(\lambda v/\sqrt{2})^2 }\simeq\frac1{\lambda^2}\left(\frac{m_{H^\pm}}{170\,{\textrm{GeV}}}\right)^2\,.
\end{equation}
While obtained in a rough limit and thus subject to potentially large correction, the above estimate is sufficient to show that Natural \mbox{$\lambda$SUSY} requires light extra scalars, to be searched for directly, or indirectly through their effect on the Higgs couplings.

Coupling modifications in \mbox{$\lambda$SUSY} have been studied in Ref.~\cite{Barbieri:2013nka} with the results reported in Fig.~\ref{fig:lSUSY}. On the plots I superimposed (roughly, since I could only use the $m_{H^\pm}$ and $\lambda$ contours on the figure) the equal-tuning lines from Eq.~(\ref{Dl}). The orange regions are excluded by the measurements while the blue ones are not theoretically accessible. The left panel shows the limits from current measurements while the right panel is a projection for the $13$~TeV LHC with $300\,{\textrm{fb}}^{-1}$, where the limits might get significantly stronger because of the improved accuracy in the determination of the Higgs coupling to $\tau$'s. Unlike in the case of the MSSM, current coupling measurements already have a significant impact on the level of Un-Naturalness of the theory and considerable improvements are expected at $13$~TeV. Notice however that the parameter space of the \mbox{$\lambda$SUSY} model is much larger to the one depicted in Fig.~\ref{fig:lSUSY}, which refers to the limiting situation where the extra singlet decouples. The case in which the singlet is light and the extra doublet decouples is also studied in Ref.~\cite{Barbieri:2013nka}, but a complete exploration of the parameter space and the systematic inclusion of the direct constraints from extra scalars searches is still missing. Because of its connection with Naturalness, I believe that this is an important subject that deserves further attention and dedicated experimental studies.

\section{Conclusions}

In this talk I tried to give (yet) another viewpoint on why the SM Higgs discovery has been a revolutionary event for the physics of fundamental interactions. By discussing No-Lose Theorems I hope I transmitted the idea that the Higgs discovery is at the same time a success and a failure for theorists. It is a success because it confirms the validity of a theory (the SM) with an amazing predictive power, which potentially extends up to the Planck scale. It is a failure for the very same reason, which makes us unable to formulate new No-Lose Theorems and to offer guarantees of new physics discoveries in the foreseeable future. Guaranteed discoveries are not common in Science and not even in Physics if a long-term historical perspective is taken. However, guaranteed discoveries have always been with us in the last half a century, to the point that we got used to them and we now lack their guidance in organising our future efforts in the exploration of fundamental interactions. On a (relatively) short time scale, the Naturalness Problem still provides such a guidance even though it does not guarantee concrete new physics discoveries. I described how the Naturalness Argument challenges the possibility of postponing to very high energies the microscopic explanation of the Higgs mass and of the EWSB scale, suggesting new physics at the TeV. The possibility of discovering Natural new physics at the $13$~TeV LHC is still open and must be explored to the best of our capabilities. The lack of discovery, {\it{i.e.}} the discovery of ``Un-Naturalness'', would still constitute a fundamental result that will stimulate future research.

Afterwards, I described the concrete impact of current LHC results on specific BSM scenarios related with Naturalness, mostly focusing on the physics associated with the scalar Higgs sector. In Composite Higgs, I outlined the importance of Higgs couplings measurements in view of their direct connection with the level of Naturalness of the theory. In SUSY, I discussed how Higgs couplings modifications and extra scalars are not directly relevant for Naturalness in the MSSM, but extremely important in non-minimal scenarios like the $\lambda$SUSY model. I also argued that after the discovery of the $125$~GeV it is not worth insisting with the MSSM as a ``Natural'' candidate model for low-energy SUSY. I also briefly discussed the future perspectives on Higgs coupling measurement. Significant improvements are expected at the $13$~TeV LHC and at the HL-LHS only in the SUSY case because of the improved determination of the Higgs coupling to $\tau$'s. A radically better sensitivity, with far-reaching implications on both the composite Higgs and the SUSY scenarios, will require future lepton colliders.

\end{document}